\documentclass[smallextended]{svjour3}       
\smartqed  
\usepackage{graphicx}

\usepackage{epstopdf}

\usepackage{xparse}
\NewDocumentCommand{\ceil}{s O{} m}{%
  \IfBooleanTF{#1} 
    {\left\lceil#3\right\rceil} 
    {#2\lceil#3#2\rceil} 
}
\NewDocumentCommand{\floor}{s O{} m}{%
  \IfBooleanTF{#1} 
    {\left\lfloor#3\right\rfloor} 
    {#2\lfloor#3#2\rfloor} 
}

\begin{document}

\title{Analogue algorithm for parallel factorization of an exponential number of large integers} 

\subtitle{II. Optical implementation}


\author{Vincenzo Tamma
}


\institute{
              Institut f\"{u}r Quantenphysik and Center for Integrated Quantum Science and Technology (IQ\textsuperscript{ST}), Universit\"{a}t Ulm, Albert-Einstein-Allee 11, D-89081 Ulm, Germany \\
              Tel.: +49 (731) 50-22781\\
              Fax: +49 (731) 50-23086\\
              \email{vincenzo.tamma@uni-ulm.de
}           
}

\date{Received: date / Accepted: date}

\maketitle

\begin{abstract}
We report a detailed analysis of the optical realization \cite{thesis,jmp,prl,foundations} of the analogue algorithm  described in the first paper of this series \cite{pra1} for the simultaneous factorization of an exponential number of integers. Such an analogue procedure, which scales exponentially in the context of first order interference, opens up the horizon to polynomial scaling by exploiting multi-particle quantum interference. 
\end{abstract}

\section{Introduction}

Factorization of a large integer $N$ is a very difficult problem to solve with our current digital computers. Indeed, divisions of $N$ for all its possible trial factors are costly tasks for a digital computer. Shor's algorithm \cite{shor,shorexp} is the only algorithm which so far would allow an exponential speed-up in the solution of the factoring problem by employing entanglement between quantum systems \cite{Lomonaco2002,nielsen}.

Recently different methods for factorization based on exponential sums \cite{primenumbers} have led to several important publications \cite{clauser,summhammer,gauss1,gauss2,gauss3,mack,merkel1,stefanak1,stefanak2,wolk,rangelov,mehring,mahesh,peng,bigourd,weber,gilowsky,sadgrove}. In the first paper \cite{pra1} of this series we have described the physical principle  for a generic analogue implementation of a novel factorization algorithm based on the analogue measurement of the periodicity in the maxima of  Continuous Truncated Exponential Sums (CTES)  \cite{thesis,jmp,prl,foundations}. 
Differently from previous factoring methods this algorithm allows the factorization of an exponential number of integers by the analogue implementation of a polynomial number of CTES interferograms where divisions of large numbers are performed by ``nature''.

Is there an example of a physical system able to compute such a factoring algorithm? 

The answer is yes! Divisions occur in a natural way in the wave nature of light. In fact, a wave of wavelength $\lambda$ propagating over a distance $d$ acquires a phase $\phi = 2 \pi d/\lambda$ and therefore naturally performs the ratio $d/\lambda$. In contrast to a digital computer, division turns out to be an instantaneous task for such a physical system.
A polychromatic source of light, which contains a continuous broad range of wavelengths, allows us to perform in parallel an exponential number of ``expensive'' divisions to test trial factors simultaneously. 

In Section II we will describe how the CTES algorithm  can be  physically implemented with an ``optical computer'' based on a polychromatic source, a multi-path Michelson interferometer and a spectrometer.  The simultaneous factorization of an exponential number of integers will be demonstrated by suitably rescaling the wavelengths of the output optical CTES interferograms. Section III will detail the experimental realizations with our optical computer of CTES of orders $j=1,2,3$ leading to the factorization of numbers with up to seven digits. Section IV and V will address, respectively, the final remarks as well as possible extensions of our optical algorithm to factoring methods with polynomial scaling.



\begin{figure}[t]
   \begin{center}
   \begin{tabular}{c}
  \includegraphics[width=1\textwidth,natwidth=610,natheight=642]{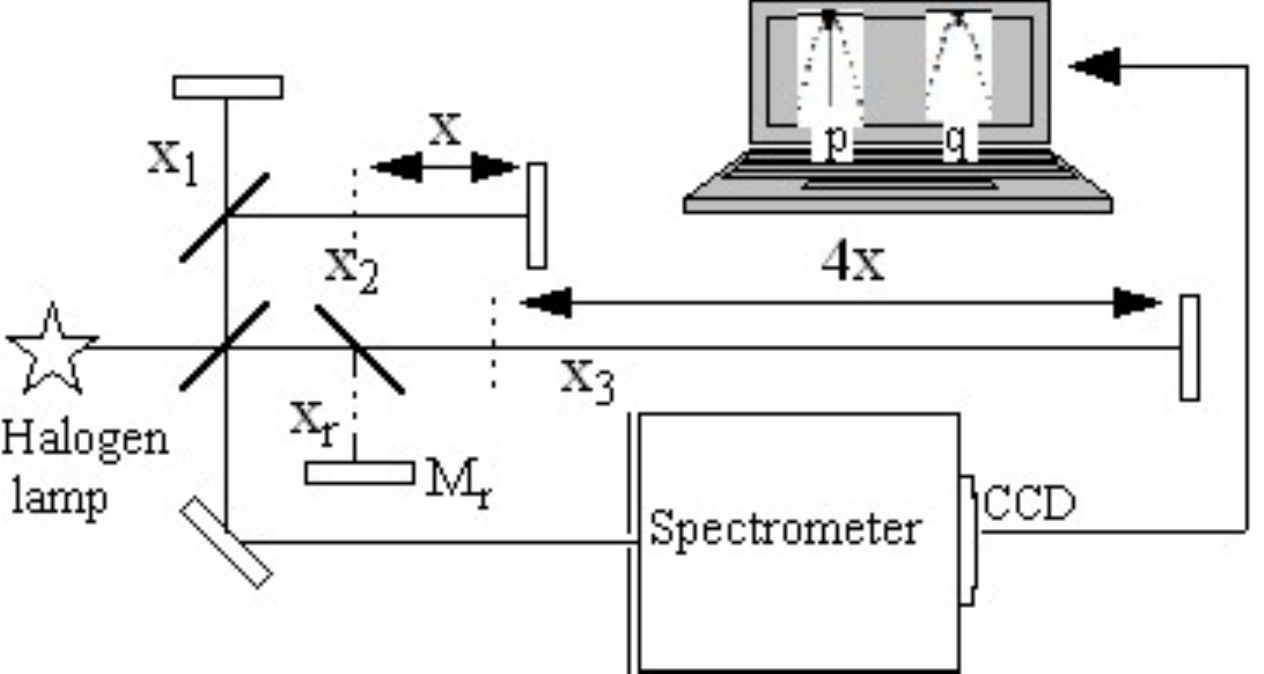}
   \end{tabular}
   \end{center}
\caption {Example with $M=3$ interfering paths of an optical computer based on a generalized symmetric Michelson interferometer, a polychromatic source (halogen lamp),  $M$ balanced beam splitters, $M+1$ mirrors, and a spectrometer connected to a CCD camera \cite {prl}. The lengths of the $M$ interfering paths are varied with respect to the reference length $x_r$ indicated by thin vertical dashed lines. The length of the $m$-th optical path reads $x_{m} = x_r + (m-1)^{j} x$, with the unit of displacement $x$, integer order $j$ (we depeict here the case $j=2$), and $m=1,2,...,M$.    \label{setup}}
\end{figure}

\section{Optical computer for factoring an exponential number of integers}
We consider a symmetric Michelson interferometer in free space with $M+1$ paths and a polychromatic source given by a halogen lamp as shown in Fig. \ref{setup} for $M=3$. The system includes $M$ balanced beam splitters and $M+1$ mirrors. 
The lengths 
\begin{eqnarray}\label{opm}
x_{m} \equiv x_r + (m-1)^j x,
\end{eqnarray}
with $m=1,2,...,M$, are calibrated with respect to the reference path $x_r$. The integer value $j$ will define the order of the CTES to be experimentally recorded. Here $x$ denotes the optical-path unit of displacement.
After the calibration, the reference mirror is blocked. The intensity in the exit port of the interferometer is the result of the interference of the waves in the remaining $M$ arms. Since we deal with a symmetric interferometer consisting of balanced beam splitters, all the interfering beams have, in principle, the same amplitude. Therefore, by normalizing the output intensity with respect to the source intensity, the different interfering optical paths $x_m$, with $m=1,2,...,M$, lead to the CTES patterns
\begin{eqnarray}\label{cesoptical}
I^{(M,j)}(\lambda;x) &\equiv & I^{(M,j)}(\displaystyle  \frac{\lambda}{x}\equiv \xi)\nonumber\\
&\equiv & \displaystyle \bigg \vert \frac{1}{M}\sum_{m=1}^{M} \exp\left[
2\pi  i (m-1)^j \displaystyle \frac{1}{\xi}\right] \bigg \vert ^2 .
\end{eqnarray}
Such interferograms, for each given value of the optical-path unit x, depend only on the dimensionless variable 
\begin{eqnarray}\label{xi}
\xi \equiv \frac{\lambda}{x}
\end{eqnarray}
given by the ratio between the wavelengths of the halogen source and the unit $x$. 



Since $x$ and $\lambda$ only enter into the spectrum as a ratio $ x/\lambda = 1/\xi$ we can easily apply the scaling law
\begin{eqnarray}
I^{(M,j)}(\lambda;x) &\equiv I^{(M,j)}(\displaystyle\frac{\lambda}{x}\equiv \xi)=I^{(M,j)}(N \frac{\lambda}{x}\equiv \xi_N;N),\nonumber\\
\end{eqnarray}\label{RescaledInterferogram}
with
$$I^{(M,j)}(\xi_N;N) = \bigg|\frac{1}{M} \sum_{m=1}^{M} \exp\left[
2\pi  i (m-1)^j \frac{N}{\xi_N}\right]\bigg|^{2}$$
continuous function of the N-dependent dimensionless variable
\begin{eqnarray}\label{xiNlambda}
\xi_N \equiv N \frac{\lambda}{x}.
\end{eqnarray}
This scaling procedure allows us to determine the factors of $N$ as the integer values  $\xi_N \equiv N \lambda/x =\ell$ whose corresponding  wavelengths $\lambda$ are associated with dominant maxima in the CTES optical interferograms in Eq. (\ref{cesoptical}).
%
%

In the next sections we will describe how this result leads to an optical implementation of the analogue algorithm described in Ref. \cite{pra1}. The two key physical observables $O_{\xi}$ and $O_x$  for such an optical computer correspond, respectively, to the source wavelengths with values $o_{\xi}\equiv \lambda$ in a given range $\lambda_{min}\leq \lambda \leq \lambda_{max}$   and the optical-path relative lengths with values $o_x^{(m,j)}\equiv x_m - x_r = (m-1)^j x$, with $m=1,2,...,M$, for given values of $M$, $j$ and path unit $x$.

\subsection{Factorization with a single optical interferogram}
In this section, we describe the CTES factoring procedure based on the measurement of a single  interferogram $I^{(M,j)}(\lambda;x)$ in Eq. (\ref{cesoptical}) recorded at a given value of $x$. We address the question of the interval $N_{min,x}\leq N\leq N_{max,x}$ of numbers $N$  factorable by covering all the (integer) trial factors $\ell$ in either the range $[3,\sqrt{N}]$  \footnote{We are sure that there is at least one factor of $N$ in such interval. The trial factor $2$ is obviously excluded since it is easy to recognize if $N$ is an even integer.} or $[\sqrt{N},N]$  with a given selected range $\lambda_{min}\leq \lambda \leq \lambda_{max}$ of wavelengths of the optical source.

 In general, for each integer $N$ a trial factor $\ell \leq N$ can be checked only if 
\begin{eqnarray}\label{lminlmaxNo}
\xi_N = \ell \in [\xi_{N}^{(min)},\xi_{N}^{(max)}] \equiv [\frac{N}{x} \lambda_{min}, \frac{N}{x} \lambda_{max}],
\end{eqnarray}
where $\xi_{N}^{(min)}$ and $\xi_{N}^{(max)}$ are respectively the smallest and largest values that the variable $\xi_N$ in Eq. (\ref{xiNlambda}) can assume for the rescaled interferogram $I^{(M,j)}(\xi_N;x)$ in Eq. ($4$).

We consider first the case in which we want to check all the  trial factors $\ell \in[3,\sqrt{N}]$   leading  from Eq. (\ref{lminlmaxNo}) to the condition
\begin{eqnarray}\label{band1N}
\mbox{Method $(1)$:} \ \xi_N = \ell \in[3,\sqrt{N}]\subseteq [\frac{N}{x} \lambda_{min}, \frac{N}{x} \lambda_{max}].
\end{eqnarray}
It is easy to obtain \cite{pra1}  the interval \footnote{We recall that for any real number $y$ the ceiling $\ceil*{y}$ is the smallest integer larger than $x$, while the floor $\floor*{y}$ is the largest integer lower than $y$.}
\begin{eqnarray}
N_{min,x}^{(1)}\equiv \ceil*{\displaystyle\frac{x^2}{\lambda_{max}^2}}\leq N \leq N_{max,x}^{(1)}\equiv \floor*{\displaystyle\frac{3 x}{\lambda_{min}}}
\end{eqnarray}
of factorable integers in the optical range $\lambda_{min}\leq \lambda \leq \lambda_{max}$ with a single interferogram associated with a given value $x$ satisfying the condition \cite{pra1}
\begin{eqnarray}\label{xlimit}
 x\leq x^{(1)}\equiv \displaystyle\frac{3 \lambda_{max}^2}{\lambda_{min}}.
\end{eqnarray}

In  the particular case of an interferogram recorded at the maximum possible value $x=x^{(1)}$ in Eq. (\ref{xlimit}), by considering the largest wavelength range such that $\lambda_{max} / \lambda_{min}$ is an integer, we find the largest but also the only integer
\begin{eqnarray}\label{Nmax1}
N^{(1)}\equiv \displaystyle  \frac{9 \lambda_{max}^2}{\lambda_{min}^2}
\end{eqnarray}
factorable by using a single experimental interferogram 
$I^{(M,j)}(\lambda;x)$.

We consider now the second case in which we want to check all the  trial factors $\xi_N = \ell \in [\sqrt{N},N]$   leading  from Eq. (\ref{lminlmaxNo}) to the condition
\begin{eqnarray}\label{band2N}
\mbox{Method $(2)$:} \ \xi_N = \ell \in[\sqrt{N}, N]\subseteq [\frac{N}{x} \lambda_{min}, \frac{N}{x} \lambda_{max}].\nonumber\\
\end{eqnarray}
In this case we easily obtain \cite{pra1} the interval  
\begin{eqnarray}
 N_{min}^{(2)}\equiv 1 \leq N \leq\floor*{\displaystyle\frac{x^2}{\lambda_{min}^2}}\equiv  N_{max,x}^{(2)}
\end{eqnarray}
of factorable numbers for the given optical range $\lambda_{min}\leq \lambda \leq \lambda_{max}$ with a single interferogram associated with a given value $x$ 
 satisfying the condition \cite{pra1}
\begin{eqnarray}\label{xlimit2}
 x\leq x^{(2)}\equiv \lambda_{max}.
\end{eqnarray}

In particular, for an optical interferogram recorded at the maximum value $x=x^{(2)}\equiv \lambda_{max}$ in Eq. (\ref{xlimit2}) we obtain the largest interval
\begin{eqnarray}\label{Nmax2}
N_{min}^{(2)}\equiv 1 \leq N \leq  \displaystyle  \floor*{\displaystyle\frac{\lambda_{max}^2}{\lambda_{min}^2}}\equiv N_{max}^{(2)}
\end{eqnarray}
of factorable integers.

We finally demonstrated that with a single optical interferogram it is possible to factorize a  number $$\Delta N \sim N_{max}\sim \displaystyle  \frac{\lambda_{max}^2}{\lambda_{min}^2}\sim 2^{n_{max}}$$ of integers exponential with respect to the number of binary digits $n_{max}$ associated with $N_{max}$.
However, in general, the largest factorable integer $N_{max}$ is limited by the value $\lambda_{max}/\lambda_{min}$ associated with the optical spectrum of the interferometer source. For this reason in the next section we will describe a factorization procedure which takes advantage of several optical interferograms  $I^{(M,j)}(\lambda;x)$  in Eq. (\ref{cesoptical}) at different values $x$ in order to factor  numbers within a limited given range of wavelengths $\lambda $. In such a method the maximum factorable number $N_{max}$ will depend on the largest value achievable for the path-unit $x$.

\subsection{Factorization with a sequence of optical interferograms}\label{sequencealgo}

So far we have restricted ourselves to a factorization method involving a single optical interferogram $I^{(M,j)}(\lambda;x)$  in Eq. (\ref{cesoptical}) defined at a fixed value of the parameter $x$. However, the remarkable scaling property $\xi_N\equiv N \lambda/x$ characterizing the function  $I^{(M,j)}(\lambda;x)$ allows us to consider not only an entire continuous range of wavelengths  $\lambda$ associated with the source but also different discrete values of the unit $x$ characterizing the optical paths in the interferometer. 

We determine how the number $n$ of experimental interferograms recorded at different values of $x$ depends on the given range $N_{min}\leq N \leq N_{max}$ of numbers to be factored.

We first point out that a single interferogram registered at a given value $x$ allows us to check for each given integer $N$  only the trial factors
\begin{eqnarray}\label{lminlmaxN}
\xi_N = \ell \in [\xi_{N}^{(min)},\xi_{N}^{(max)}] \equiv [\frac{N}{x} \lambda_{min}, \frac{N}{x} \lambda_{max}].
\end{eqnarray}
In general, for the fixed domain $\lambda_{min}\leq \lambda\leq \lambda_{max}$ of values $\lambda$, these trial factors may correspond only to a subset of the total range $3\leq \ell \leq \sqrt{N}$ or $\sqrt{N}\leq \ell \leq N$ of integer values $\xi_N=\ell$ we would need to cover in order to factor a generic integer $N$. For this reason, we consider a suitable sequence of $n$ (to be determined) values  $x=x_i$, with $i=0,1,...,n-1$. Each interferogram registered at the value  $x=x_i$, with $i=0,1,...,n-1$, allows us from Eq. (\ref{lminlmaxN}) to cover all the trial factors 
\begin{eqnarray}\label{liN}
\xi_N = \ell \in [\xi_{N,i},\xi_{N,i+1}] \equiv [\frac{N}{x_{i}}\lambda_{min},  \frac{N}{x_{i}}\lambda_{max}],
\end{eqnarray}
with $i=0,1,...,n-1$, where
\begin{eqnarray}
\xi_{N,i+1}\equiv \frac{N}{x_{i}}\lambda_{max}\equiv \frac{N}{x_{i+1}}\lambda_{min}
\end{eqnarray}
satisfies the condition for consecutive intervals. This implies that the sequence $x_i$, with $i=0,...,n-1$, defining the $n$ interferograms to be recorded, follows the iterative formula
\begin{eqnarray}\label{x_iN}
x_{i+1}\equiv\frac{x_i}{c}  < x_i,
\end{eqnarray}
with
\begin{eqnarray}\label{c2}
c\equiv \frac{\xi_{N,i+1}}{\xi_{N,i}}=\frac{\lambda_{max}}{\lambda_{min}}>1.
\end{eqnarray}
%
In particular, factorization can be achieved for all values 
of  $N$, with $N_{min}\leq N \leq N_{max}$,  only if for each single value is satisfied either the condition   
\begin{eqnarray}\label{range1}
&\mbox{Method $(1)$:} \\ &\xi_N = \ell \in [3,\sqrt{N}]\subseteq [\xi_{N,0},\xi_{N,n}]\equiv [\displaystyle\frac{N \lambda_{min}}{x_{0}},\frac{N \lambda_{min}}{x_{0}} \ c^n]  \nonumber
\end{eqnarray}
or the condition  
\begin{eqnarray}\label{range2}
&\mbox{Method $(2)$:} \\ &\xi_N = \ell \in[\sqrt{N},N]\subseteq [\xi_{N,0},\xi_{N,n}] \equiv \displaystyle[\frac{N \lambda_{min}}{x_{0}},\frac{N \lambda_{min}}{x_{0}} \ c^n] \nonumber
\end{eqnarray}
for the total interval $[\xi_{N,0},\xi_{N,n}]$ of  trial factors covered by the $n$ interferograms.

In the factorization method $(1)$ the interferograms $I^{(M,j)}(\lambda;x)$ are measured at the values $x=x_i$, with $i=0,1,...,n-1$, defined by Eq. (\ref{x_iN}) with
\begin{eqnarray}\label{x01}
\frac{x_{0}}{\lambda_{min}} \geq \frac{x_{0}^{(1)}}{\lambda_{min}} \equiv \frac{N_{max}}{3}.
\end{eqnarray}
From Ref. \cite{pra1} we also obtain the minimum number 
\begin{eqnarray}\label{n1}
n_{x_0}^{(1)} &\equiv & \ceil* {log_{c} \ \frac{x_{0}}{\lambda_{min}\sqrt{N_{min}}}} \nonumber\\
&\geq &\ceil*{log_{c}\ \frac{N_{max}}{3\sqrt{N_{min}}}}\equiv n^{(1)}_{min}
\end{eqnarray}
of  interferograms  necessary to factor all the integers $N$ in any given interval  $N_{min}\leq N \leq N_{max}$.

We consider now the method $(2)$ associated with the condition in Eq. (\ref{range2}).
In such a case, the interferograms $I^{(M,j)}(\lambda;x)$ are recorded at the values $x=x_i$, with $i=0,1,...,n-1$, defined by Eq. (\ref{x_iN}) with
\begin{eqnarray}\label{x02}
\frac{x_{0}}{\lambda_{min}} \geq \frac{x_{0}^{(2)}}{\lambda_{min}}\equiv \sqrt{N_{max}}.
\end{eqnarray}
We also easily obtain \cite{pra1} the minimum number
\begin{eqnarray}\label{n2}
n^{(2)}_{x_0} &\equiv &\ceil*{log_{c}\frac{x_{0}}{\lambda_{min}}}
\geq  \ceil*{log_{c} \sqrt{N_{max}}}\equiv n_{min}^{(2)}
\end{eqnarray} 
of  interferograms necessary to factor all the integers $N$ in any given interval  $N_{min}\leq N \leq N_{max}$.

We have finally demonstrated that the number $n$ of experimental runs necessary for factorizing any given range of numbers $N_{min}\leq N \leq N_{max}$, with $N_{min} \geq 1$, using a selected wavelength spectrum $[\lambda_{min},\lambda_{max}]$, scales logarithmically with respect to either  $N_{max}/\sqrt{N_{min}}$ (method (1)) or $\sqrt{N_{max}}$ (method (2)). The described algorithm allows the factorization of a number
$$\Delta N \sim N_{max}\sim 2^{n_{max}}$$
 of integers exponential with respect  to the number $n_{max}$ of binary digits of $N_{max}$  by using a polynomial number of interferograms if $$ c \equiv \frac{\lambda_{max}}{\lambda_{min}}\geq 2 .$$ On the other hand, the  parameter $x_0$ for the first interferometer scales exponentially with respect with $n_{max}$.

\section{Experimental realizations}

We now turn to the experimental implementation of our factoring technique. The  experimental setup consists of a symmetric Michelson interferometer with $M$ interfering paths of which we have given an example for $M=3$ in Fig. $1$.   Each mirror is mounted on a single axis translation stage.  Each stage consists of a $5 mm$ manual travel stage, a $50 mm$ step motor with $58200$ steps for each $mm$, and a $20 \mu m$ piezoelectric and feedback control stage. The resolution of the piezoelectric element is $10 \, nm$.  The polychromatic source of the interferometer is given by a halogen lamp while a He-Ne laser expanded by lenses is used for the alignment. The interference pattern  at the output port of the interferometer is measured by a spectrometer as a continuous function of the wavelengths $\lambda$ associated with the polychromatic source. In the experiments we will consider the visible spectrum. The spectrometer, with resolution $0.01 \, nm$, is characterized by a grating composed by $2400$ elements for each $mm$ and by a $2048$-pixel CCD array with an accuracy of $0.005-0.006 \, nm$.

Calibrating the optical paths given by Eq. (\ref{opm}) with a suitable accuracy is one of the most challenging tasks in this experiment. We first determine when all path lengths $x_m$ are equal to $x_r$, by measuring the polychromatic two-path interference between the $m^{th}$ beam and the reference beam, for each $m=1,2,...,M$, with the mirror $M_{r}$ tilted by a small angle with respect to all the other mirrors. In particular, the interference fringe associated with two equal paths is completely bright \footnote{The fringe can also be completely dark if an extra $\pi$ shift emerges from the number of beam splitters and mirrors present in the two paths.} for all the wavelengths of the polychromatic source (``white light condition''). We calibrate the $m^{th}$ path until such a fringe is in correspondence to the entrance slit of the spectrometer. Then we block the mirror $M_r$ and translate each mirror $M_{m}$, using the piezoelectric translators together with the step motors, so that  Eq. (\ref{opm}) is satisfied. The optical interferometer is now prepared to record the CTES factoring interferogram in Eq. (\ref{cesoptical}) for the chosen values of $x,j$ and $M$.

In the next sections we will show experimental demonstrations of our factoring method with CTES interferograms of different orders $j=1,2,3$ and numbers of interfering optical paths $M=2,3$. We will show how it is possible to select factors of different numbers for a given experimental interferogram recorded at a particular value of the path unit $x$. Such demonstrations can be easily extended for the generic implementation of the factoring algorithm described in section \ref{sequencealgo} which allows us to check all the possible trial factors of any integer less than $N_{max}$ with only a polynomial number of CTES interferograms recorded, for example, in the optical range $400 \, nm \leq \lambda \leq 800 \, nm$ ($\lambda_{max}/\lambda_{min}=2$).

\subsection{Experimental results for $j=2$ and $M=3$}

After having described the experimental preparation, we can focus now on the actual measurement of the CTES optical interferogram $I^{(M,j)}(\lambda;x)$ in Eq. (\ref{cesoptical}). In this section, we will consider the case of $M=3$ interfering phase terms with a quadratic dependence ($j=2$) on $m$ leading to Continuous Truncated Gauss Sum (CTGS) interferograms.


\begin{figure*}[h]
   \begin{center}
   \begin{tabular}{c}
  \includegraphics[width=1\textwidth,natwidth=610,natheight=642, angle=270]
   {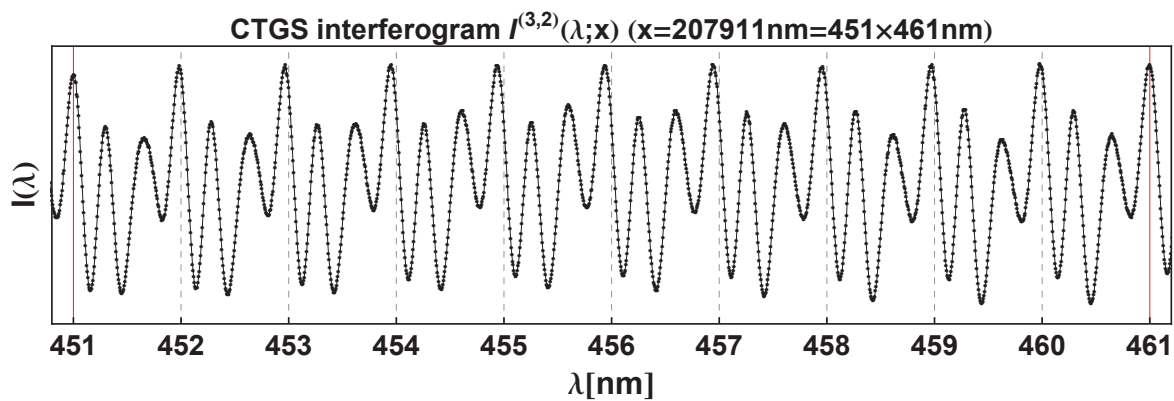}
   \end{tabular}
   \end{center}
\begin{quote}\caption [Experimental realization of the CTGS  optical interferogram of $M=3$ interfering paths with unit of displacement $x=207911 \, nm$ for the factorization of $N=x/nm$.]{Experimental realization of the CTGS ($j=2$) interferogram $I(\lambda)=I^{(M,j)}(\lambda;x)$ in Eq. (\ref{cesoptical}) for $M=3$ and unit of displacement $x=207911 \, nm$, in the wavelength range $450.173\, nm \leq \lambda \leq 461.934\, nm$ \cite{thesis}.  The dots represent the measured values, and the curve is obtained by joining these experimental points. Only the trial factors $\ell=451$ and $\ell=461$, represented by continuous vertical lines,  correspond  to the brightest integer wavelengths and therefore are the factors of $N$. Instead,  all the other integer wavelengths, represented by dashed vertical  lines, are not associated with local  maxima and consequently are not factors.\label{experimentalresultsN=451X461}}
\end{quote}\end{figure*}


In Fig. \ref{experimentalresultsN=451X461} we give an experimental proof of principle for the factorization of $N=207911$. The recorded CTGS interferogram measured for the unit of displacement $x=N \, nm$ in the wavelength range $450.173 \, nm\leq \lambda \leq 461.934 \, nm$ is scaled according to the corresponding  auxiliary variable $\xi_N=\lambda/nm$ in Eq. (\ref{xiNlambda}). Thereby, the integer values of $\xi_N$ are also integer values of wavelengths measured in $nm$.   All the trial factors $\xi_N=\ell$ of $N$ corresponding to the  wavelengths $\lambda =  \ell \, nm$ are represented by vertical lines.
 Only the trial factors $\ell=451$ and $\ell=461$, represented by continuous lines,  correspond  to the brightest integer wavelengths and therefore are the factors of $N$. Instead,  all the other integer wavelengths, represented by dashed lines, are not associated with dominant local maxima and consequently are not factors.
It is interesting to note that in the wavelength range in Fig. \ref{experimentalresultsN=451X461} the closer a non factor is to a factor, the higher is the  corresponding value of intensity. Therefore, we can expect that when non factors are far away from factors, it is generally much easier to recognize that the associated wavelengths do not correspond to dominant local maxima,
as we have shown in Ref. \cite{pra1}.  In conclusion, we have demonstrated that the factors of $N=207911$
correspond to the dominant maxima at integer wavelengths of the recorded interferogram.

\begin{figure*}[hb]
      \begin{center}
      \begin{tabular}{c}
   \includegraphics[width=1\textwidth,natwidth=610,natheight=642, angle=270]{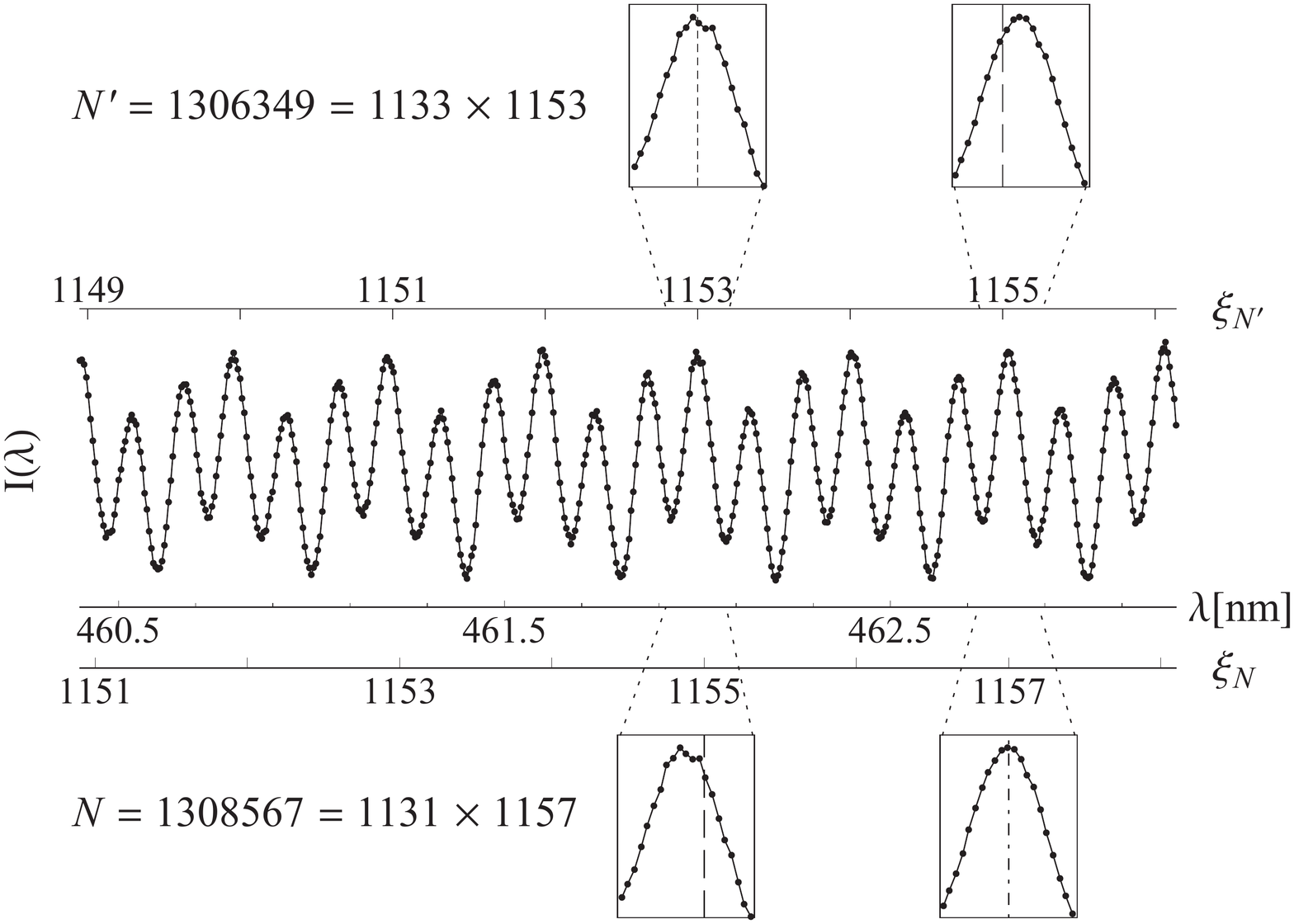}
   \end{tabular}
   \end{center}
\begin{quote}\caption {Experimental realization of the CTGS ($j=2$) interferogram $I(\lambda)=I^{(M,j)}(\lambda;x)$ in Eq. (\ref{cesoptical}) for $M=3$ and unit of displacement $x=523426.8 \, nm$, in the wavelength range $460.36 \, nm\leq \lambda \leq  463.24 \, nm$ (center) \cite{thesis,prl}. The factorization of the two numbers $N=1308567=1131\times 1157$ (bottom) and $N'=1306349=1133\times1153$ (top) is obtained by rescaling the wavelength axis according to Eq. (\ref{xiNlambda}). The insets magnify the behavior of the interferogram in the neighborhoods of the two dominant maxima corresponding to the trial factors $1153$ ,$1155$ (this integer is checked as a trial factor for both $N$ and $N'$) and $1157$,   which are indicated by dotted,  dashed, and dashed-dotted lines,  respectively. The first dominant maximum in the interferogram points to a factor ($\ell=1153$) of $N'$ but not of $N$. On the other hand, the second dominant maxima corresponds to a factor ($\ell=1157$) of $N$ but not of $N'$.\label{experimentalresults}}
\end{quote}\end{figure*}

An experimental proof of principle for the factorization of multiple numbers is given in  Fig. \ref{experimentalresults} where the CTGS optical interferogram is measured in the wavelength interval $[451.784\, nm,463.522\, nm]$,  using this time the displacement unit $x=523426.8 \, nm$.
We give an example for the factorization of two numbers, $N=1308567=1131\times 1157$ and $N'=1306349=1133\times1153$. We only consider the wavelength interval $[460.36 \, nm, 463.24 \, nm]$ shown in the center of Fig. \ref{experimentalresults}.
We start with the seven-digit number $N=1308567=1131\times 1157$ and  present on the bottom of Fig. \ref{experimentalresults} an axis with the variable $\xi_{N}$ rescaled according to Eq. (\ref{xiNlambda}). We clearly identify the factor $1157$ by the maximum being located at an integer, as shown by the inset. Moreover, we use the same interferogram to factor the number $N'=1306349=1133\times1153$. For this purpose we show  on the top  the rescaled variable $\xi_{N'}$. Again we can identify the factor $1153$ by the maximum being located at an integer. This demonstrates that the accuracy in our experiment allows us to factor numbers with values up to $N_{max}\sim 10^6$
by tilting the wavelength axis in order to obtain the auxiliary variable $\xi_N$ associated with a generic number $N<N_{max}$ to factor.

We have demonstrated that the CTES algorithm is optically computable for different values of the unit of displacement $x$. As expected, the obtained results point out that the maximum number factorable with the recorded interferogram increases with the value of the unit $x$. The upper limit for the value of $x$ is determined by the coherence length associated with the experimental conditions.

\subsection{Experimental results for $j=2$ and $M=2$}
\begin{figure*}[hb]
   \begin{center}
   \begin{tabular}{c}
   \includegraphics[width=1\textwidth,natwidth=610,natheight=642, angle=270]{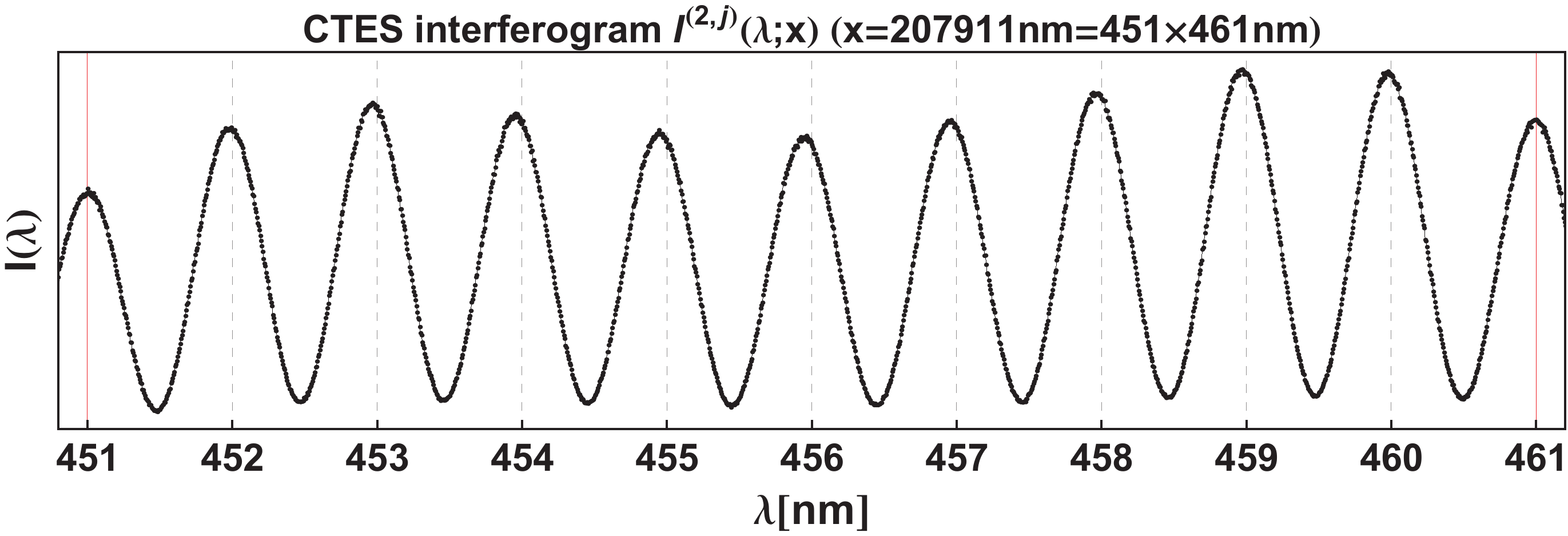}
   \end{tabular}
   \end{center}
\begin{quote}\caption [Experimental realization of the CTES optical interferogram of $M=2$ interfering paths with unit of displacement $x=207911 \, nm$ for the factorization of $N=x/nm$.]{Experimental realization of the CTES interferogram $I(\lambda)=I^{(M,j)}(\lambda;x)$ in Eq. (\ref{cesoptical}) for $M=2$, $x=207911 \, nm$, in the wavelength range $450.173 \, nm \leq \lambda \leq 461.934 \, nm$ \cite{thesis}. In this case, since there are only two interfering terms associated with $m=1,2$, the value of the order $j$ is not significant  anymore.  Only the trial factors $\ell=451$ and $\ell=461$, represented by continuous vertical lines,  correspond  to the  dominant maxima and therefore are the factors of $N$. Instead,  all the other integer wavelengths, represented by dashed vertical lines, are not associated with local  maxima and consequently are not factors.\label{experimentalresultsN=451X461M=2}}
\end{quote}\end{figure*}

We now consider the case of the CTES optical interferogram $I^{(M,j)}(\lambda;x)$ in Eq. (\ref{cesoptical}) for $M=2$ and unit of displacement $x=N \, nm$, with $N=207911$. In this case, since we have only two interfering terms associated with $m=1,2$, the value of the order $j$ is not significant anymore. We have experimentally recorded such an interferogram in the wavelength range $450.173 \leq \lambda \leq 461.934$ (see Fig. \ref{experimentalresultsN=451X461M=2}). Again we can distinguish the factors $451$ and $461$ as the brightest integer wavelengths in the pattern.

It is interesting to compare the obtained interferogram for $M=2$ interfering terms in Fig. $4$ with the one for $M=3$ terms in Fig. \ref{experimentalresultsN=451X461}. In the case of $M=2$ the secondary peaks disappear and the dominant peaks are wider \cite{pra1}. 


\subsection{Experimental results for $j=1,3$ and $M=3$}

\begin{figure*}[hb]
\centering
   \begin{tabular}{c}
   \includegraphics[width=1\textwidth,natwidth=610,natheight=642, angle=270]{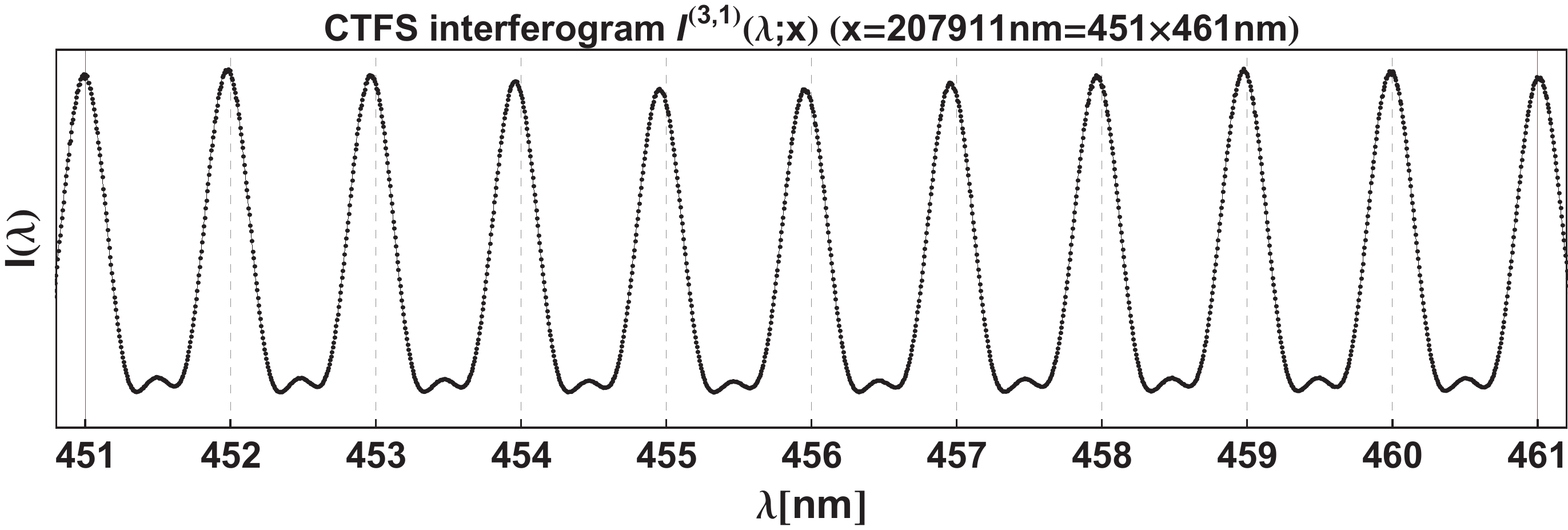}
   \end{tabular}
\begin{quote}\caption [Experimental realization of the CTFS  optical interferogram of  $M=3$ interfering paths with unit of displacement $x=207911 \, nm$ for the factorization of $N=x/nm$.]{Experimental realization of the CTFS ($j=1$) interferogram $I(\lambda)=I^{(M,j)}(\lambda;x)$ in Eq. (\ref{cesoptical}) for $M=3$ and  $x=207911 \, nm$, in the wavelength range $450.173 \, nm \leq \lambda \leq 461.934 \, nm$ \cite{thesis}. We recognize the factors $\ell=451$ and $\ell=461$ of $N$, represented by continuous vertical lines, as the dominant maxima with respect to the other trial factors,  represented by dashed vertical lines.\label{experimentalresultsN=451X461j=1}}
\end{quote}\end{figure*}

\begin{figure*}[hb]
   \begin{center}
   \begin{tabular}{c}
   \includegraphics[width=1\textwidth,natwidth=610,natheight=642, angle=270]{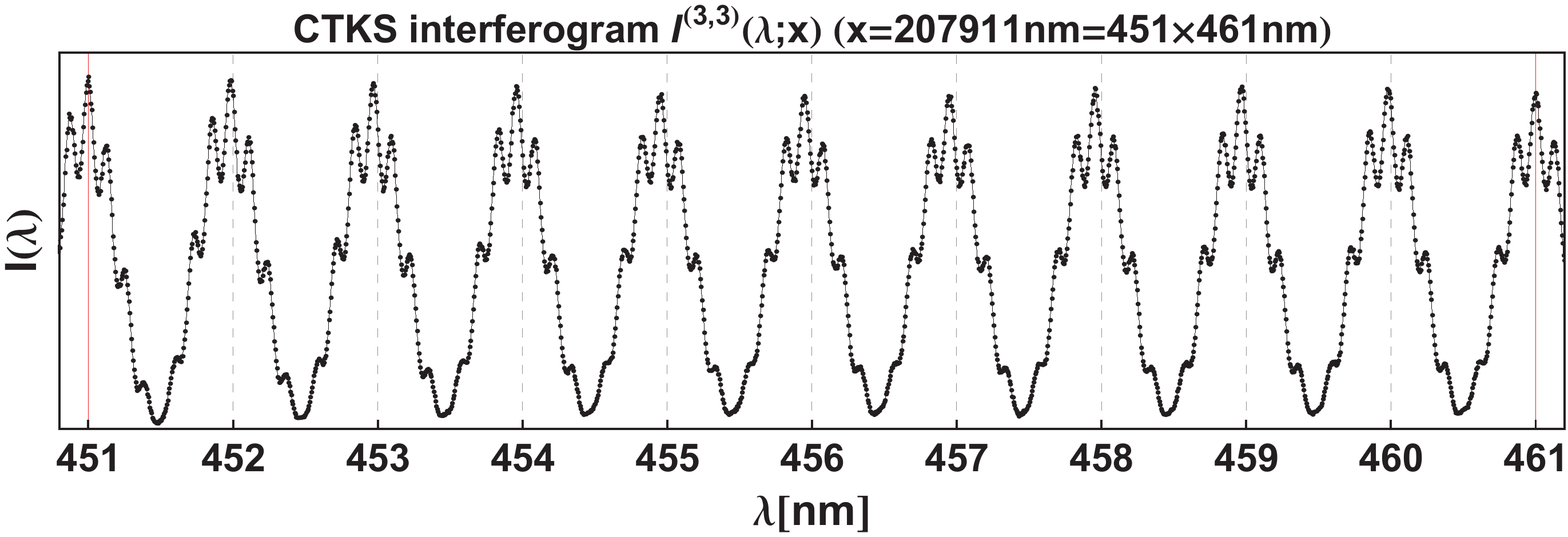}
   \end{tabular}
   \end{center}
\begin{quote}\caption  [Experimental realization of the CTKS optical interferogram of $M=3$ interfering paths with unit of displacement $x=207911 \, nm$ for the factorization of $N=x/nm$.]{Experimental realization of the CTKS ($j=3$) interferogram $I(\lambda)=I^{(M,j)}(\lambda;x)$ in Eq. (\ref{cesoptical}) for $M=3$ and $x=207911 \, nm$, in the wavelength range $450.173 \, nm \leq \lambda \leq 461.934 \, nm$ \cite{thesis}. Only the trial factors $\ell=451$ and $\ell=461$, represented by continuous vertical lines,  correspond  to the brightest integer wavelengths and therefore are the factors of $N$. Instead,  all the other integer wavelengths, represented by dashed vertical lines, are not associated with local  maxima and consequently are not factors. \label{experimentalresultsN=451X461j=3}}
\end{quote}\end{figure*}

We now consider the case of $M=3$ interfering phase terms and $x=207911 \, nm$ for two different orders $j=1,3$ of the CTES.  In Fig. \ref{experimentalresultsN=451X461j=1} is represented the recorded interferogram for $j=1$ corresponding to a Continuous Truncated Fourier Sum (CTFS). Instead, in Fig. \ref{experimentalresultsN=451X461j=3} is shown the measured Continuous Truncated Kummer Sum (CTKS) interferogram corresponding to $j=3$.  In both pattern we can recognize the factors $451$ and $461$ as the brightest integer wavelengths.

Comparing the interferograms in Figs. \ref{experimentalresultsN=451X461j=1}, \ref{experimentalresultsN=451X461}, \ref{experimentalresultsN=451X461j=3}, we can note that  peaks of higher order appear as the order $j$ increases and at the same time the dominant peaks important for factorization become sharper \cite{pra1}.



\section{Remarks}

We have demonstrated the physical computability of the CTES algorithm using an optical computer characterized by a multi-path Michelson interferometer, a spectrometer and a polychromatic optical source.

Such an optical computer exploits destructive/constructive interference to experimentally compute the CTES optical interferograms  $$I^{(M,j)}(\lambda;x)\equiv I^{(M,j)}(\frac{\lambda}{x}\equiv \xi)$$ in Eq. (\ref{cesoptical}) recorded over the continuous range of wavelengths $\lambda \equiv \xi x$ of the polychromatic source, with $x$ unit of displacement in the optical paths. The wave nature of light allows us to experimentally compute the divisions  $f(1/\xi)= x/\lambda$ for all the possible wavelengths $\lambda = \xi x$ in the CTES optical spectrum. The information about such divisions can be extracted by measuring the  periodicity in the maxima of the recorded interferogram. Moreover,  rescaling such a periodicity according to the relation $\xi_N\equiv N \lambda/x$ allows us to infer information about the divisions $f(\xi_N)= N/\xi_N$ and, thereby, about the factors of several numbers $N$. Indeed, for each value of $N$, the factors are the integer values of $\xi_N\equiv N \lambda/x$ corresponding to dominant maxima of the recorded CTES interferogram.

Furthermore, we have demonstrated that an optical computer can implement prime number decomposition of an exponential number of integers $N_{min}\leq N \leq N_{max}$. In particular, the CTES factorization algorithm takes advantage of a sequence of optical interferograms $I^{(M,j)}(\lambda;x_i)$, with $i=0,1,...,n-1$, where each one is associated with a different value $x=x_i = (\lambda_{min}/\lambda_{max})\, x_{i-1}$  of the unit of displacement in the multi-path interferometer. The number $n$ of interferograms increases polynomially with respect to the logarithm in base $\lambda_{max}/\lambda_{min}$ of the largest number $N_{max}$ to be factored. Therefore, even by exploiting a source with a wavelength bandwidth only in the visible range such that $\lambda_{max}/\lambda_{min}=2$, it is possible to achieve a logarithmic scaling in base $2$.

Finally, we have given a proof of principle demonstration of the physical computability of the CTES algorithm in the visible range. The factors of  numbers with up to seven digits were experimentally found by using optical CTES inteferograms of order $j=1,2,3$ with $M=2,3$ interfering paths. We have shown, as expected, that by increasing the number $M$ of interfering terms and the order $j$ of the CTES interferogram  the dominant peaks become sharper and sharper. This property can be exploited in order to better distinguish factors from non factors.


Our experimental results demonstrate that the CTES  algorithm is not just an abstract mathematical tool but is implementable using an optical computer, which exploits the connection between physical phenomena of light interference and number theory for parallel factorization of an exponential number of integers.

\section{Towards factoring with polynomial scaling}


The described optical algorithm, in contrast to Shor' s method, leads to the factorization of an exponential number of integers by experimentally implementing a single sequence of a polynomial number of interferograms. However, our scheme does not allow the achievement of polynomial scaling in the number of resources as in the Shor's algorithm, which instead takes advantage of entanglement between  single-photon qubits in order to achieve such a goal. 
Indeed, the largest number factorable $N_{max}$ is upper limited either by the value $(\lambda_{max}/\lambda_{min})^2$ or $x_0/\lambda_{min}$, depending on the use of a single interferogram or a sequence of interferograms. Moreover, the accuracy  in the variable $\xi$ in Eq. (\ref{xi})
$$\Delta \xi = \frac{\lambda}{x^2} \Delta x + \frac{1}{x} \Delta \lambda \leq  \frac{\lambda_{max}}{x^2} \Delta x + \frac{1}{x} \Delta \lambda$$
associated with the trial factors of $N$ depends on the experimental uncertainties $\Delta \lambda$ and $\Delta x$ associated with the measurement of the wavelengths  $\lambda$ and the optical path unit $x$, respectively. The uncertainty $\Delta \xi$ also has to decrease exponentially \cite{pra1}, implying that the unit $x$  defining the CTES interferograms in Eq. (\ref{cesoptical}) has to grow exponentially.   Interferometric configurations based on multiple path-reflections may be used to achieve larger values of $x$.


An extension of the presented algorithm based on correlation measurements in $n^{th}$ order  interferometers may pave the way towards a new algorithm using a  polynomial number of resources, thereby avoiding the requirement of exponentially large values for the  optical-path unit $x$.  In this case, multi-photon quantum interference may serve as a powerful tool to distinguish factors from non factors. 

\begin{acknowledgements}
We thank H. Zhang, X. He, and Y.H. Shih for their important experimental contributions to this work and  J. Franson, M. Freyberger, A. Garuccio, S. Lomonaco, R. Meyers, T. Pittman, M. H. Rubin, W. P. Schleich for many fruitful discussions.
\end{acknowledgements}

\end{document}